\documentclass[10pt]{article}
\usepackage{hyperref}
\begin{document}
\title{Experimental Investigation of Shock Wave Surfing}
\author{N. J. Parziale$^{\dagger}$, S. J. Laurence$^{\ddagger}$, R Deiterding$^{\mp}$\\
H. G. Hornung$^{\dagger}$, J. E. Shepherd$^{\dagger}$ \\
\\ California Insitute of Technology, Pasadena, CA 91125, USA$^{\dagger}$
\\ German Aerospace Center, G\"{o}ttingen D 37073, Germany$^{\ddagger}$
\\ Oak Ridge National Laboratory, Oak Ridge, TN 37831, USA$^{\mp}$
}
\maketitle
%% The abstract (in this file, and that submitted as text to arXiv) should include the exact phrase
%% "fluid dynamics video" or "fluid dynamics videos"
Shock wave surfing is investigated experimentally in GALCIT's Mach 4.0 Ludwieg Tube. Shock wave surfing occurs when a secondary free-body follows the bow shock formed by a primary free-body; an example of shock wave surfing occurs during meteorite breakup. The free-bodies in the current investigation are nylon spheres. During each run in the Ludwieg tube a high speed camera is used to capture a series of schlieren images; edge tracking software is used to measure the position of each sphere. Velocity and acceleration are had from processing the position data. The radius ratio and initial orientation of the two spheres are varied in the test matrix. The variation of sphere radius ratio and initial angle between the centers of gravity are shown to have a significant effect on the dynamics of the system.
% main text

The air flow in each fluid dynamics video is from left to right. The Mach number is 4.0 with total pressure in each run of roughly 200kPa. Each run begins with the spheres tethered and no flow. Flow arrives and the tethers are quickly broken. The spheres are then free-bodies with the dominant force being the pressure distribution around the surface of the sphere, which, in the case of the secondary sphere, is strongly manipulated by the shock wave interaction. Gravity is not a considerable body force because of the short test times in these experiments. The secondary spheres are visibly seen to be affected by the bow shock wave of the primary bodies in the form of quite beautiful trajectories. Computational images were obtained from a simulation using the AMROC code. A fully three-dimensional simulation of two free-flying bodies was performed, assuming inviscid flow, whereby the trajectory of each sphere was calculated according to the instantaneous aerodynamic forces experienced.
\\
\small{
\begin{itemize}
\item \href{http://www.galcit.caltech.edu/~np/forStuart/forAPS2010/apsParzialeLaurence.mp4}{Shock wave surfing video uploaded to APS}
\item \href{http://www.galcit.caltech.edu/~np/forStuart/forAPS2010/apsParzialeLaurenceShot286.mp4}{Another example of shock surfing}
\item \href{http://www.galcit.caltech.edu/~np/forStuart/forAPS2010/apsParzialeLaurenceShot270.mp4}{Symmetric shock surfing}
\item \href{http://www.galcit.caltech.edu/~np/forStuart/forAPS2010/apsParzialeLaurenceShot285.mp4}{Zoomed in movie of the startup process}
\end{itemize}
}
\end{document}